\documentclass[useAMS,usenatbib]{mn2e}
\usepackage{times}
\usepackage{graphicx}

\newcommand{\kms}   {km~s$^{-1}$}

\newcommand{\eg}    {e.\,g.}
\newcommand{\ie}    {i.\,e.}

\newcommand{\sii}   {[S\,{\sc ii}]}
\newcommand{\nii}   {[N\,{\sc ii}]}

\newcommand{\vlsr}  {$V_\rmn{LSR}$}
\newcommand{\vtot}  {$V_\rmn{tot}$}

\title[Integral Field Spectroscopy of HH 110]
{The complex structure of HH 110 as revealed 
 from Integral Field Spectroscopy}

\author[R. L\'opez et al.]
{R. L\'opez,$^{1}$\thanks{E-mail:
\mbox{rosario@am.ub.es;
bgarcia@iac.es };
sanchez@cefca.es;
\mbox{gabriel.gomez@gtc.iac.es;
robert.estalella@am.ub.es;
angels.riera@upc.edu}
}
B. Garc\'\i a-Lorenzo $^{2,3}$\footnotemark[1]
S. F. S\'anchez,$^{4,5,6}$\footnotemark[1] 
G. G\'omez,$^{7,2}$\footnotemark[1]
R. Estalella,$^{1}$\footnotemark[1]
and
\newauthor 
A. Riera$^{8}$\footnotemark[1] 
\thanks{Based on observations collected at the Centro Astron\'omico Hispano
Alem\'an (CAHA) at Calar Alto, operated jointly by the Max-Planck Institut f\"ur
Astronomie and the Instituto de Astrof\'{\i}sica de Andaluc\'{\i}a (CSIC) 
}\\
$^{1}$Departament d'Astronomia i Meteorologia (IEEC-UB), Institut de Ci\`encias
del Cosmos, Universitat de Barcelona, Mart\'{\i} i Franqu\`es 1, 
E-08028 Barcelona, Spain\\
$^{2}$Instituto de Astrof\'{\i}sica de Canarias, E-38200 La Laguna, Spain\\
$^{3}$Departamento de Astrof\'{\i}sica, Universidad de La Laguna, E-38205,
 Tenerife, Spain\\
$^{4}$Centro de Estudios de F\'{\i}sica del Cosmos de Arag\'on
(CEFCA), C/General  Pizarro 1, E-41001 Teruel, Spain.\\
$^{5}$Fundaci\'on Agencia Aragonesa para la Investigaci\'on y
el Desarrollo (ARAID).\\
$^{6}$Centro Astron\'omico Hispano-Alem\'an, Calar Alto,
(CSIC-MPG), C/Jes\'us Durb\'an Rem\'on 2-2, E-04004 Almeria, Spain.\\
$^{7}$GTC Project Office, GRANTECAN S.A. (CALP), E-38712 Bre\~na Baja, La Palma,
Spain.\\
$^{8}$Dept.\ F\'{\i}sica i Enginyeria Nuclear. EUETI de Barcelona.
Universitat Polit\`ecnica de Catalunya. Comte d'Urgell 187, E-08036 Barcelona,
Spain
}

\begin{document}

\date{Accepted 2010 April 11. Received 2010 March 19; in original form 2009
December 29 }
\pagerange{\pageref{firstpage}--\pageref{lastpage}} \pubyear{2009}

\maketitle

\label{firstpage}

\begin{abstract}

HH~110 is a rather peculiar Herbig-Haro object in Orion that originates due to
the deflection of another jet (HH~270) by a dense molecular clump, instead of
being directly ejected from a young stellar object. Here we present new results
on the kinematics and physical conditions of HH~110 based on Integral Field
Spectroscopy. The 3D spectral data cover  the whole outflow extent
($\sim$~4.5~arcmin, $\simeq$ 0.6 pc at a distance of 460~pc) in the spectral
range 6500--7000 \AA.  We built emission-line intensity maps of H$\alpha$, \nii\
and \sii\ and of  their radial velocity channels. Furthermore, we analysed the
spatial distribution of the excitation and electron density  from
\nii/H$\alpha$, \sii/H$\alpha$, and \sii 6716/6731 integrated line-ratio maps,
as well as  their behaviour as a function of velocity, from line-ratio channel
maps. Our results  fully reproduce the morphology and kinematics obtained from 
previous imaging and long-slit data. In addition, the IFS data revealed, for
the first time, the complex spatial distribution of the physical conditions
(excitation and density) in the whole jet, and their behaviour as a
function of the kinematics. The results here derived give further support to the
more recent model simulations that involve deflection of a pulsed jet
propagating in an inhomogeneous ambient medium. The IFS data  give  richer 
information than that provided by  current model simulations  or laboratory jet
experiments. Hence, they could provide valuable clues to constrain the space
parameters in future theoretical works.

\end{abstract}


\section{Introduction}

HH~110 is a Herbig-Haro (HH) jet emerging from the southern edge of the L1617
dark cloud in the Orion B complex. Both observational and theoretical works have been carried out since its discovery by 
 \citet{Rei91}, because unlike most of the well-known stellar jets, HH~110 has 
a peculiar morphology, among other properties.
The morphology of HH~110 in the H$\alpha$ and \sii\ lines is very complex,
starting in a collimated chain of knots.  The emission away
from the collimated region has a more chaotic structure and widens within a
cone of unusually large opening angle ($\sim 10^\circ$). HH~110 
appreciable wiggles along the $\sim$ 4~arcmin jet length. 
The knots are all embedded in  fainter emitting gas, which
outlines the whole flow, more reminiscent of a turbulent outflow. Most of the 
knots detected in ground-based images are spatially resolved into several
components in the higher spatial resolution HST images (see \eg\
\citealp{Har09}). 
The transverse
cross-section of HH~110 shows a significant asymmetry, the eastern border is
sharp and poorly resolved, whereas the strong knotty emission mostly appears towards the
western side. HH~110 is the only known HH that shows a faint filament of
emission lying parallel at $\sim$ 10 arcsec to the east of the northern A-C 
knots, which
probably represents a weak secondary or even a fossil flow channel \citep{Rei91}.
Attempts to find the 
driving source of
HH~110 have failed at optical, near infrared and radio continuum wavelengths.

A scenario that accounts for the singular morphology of HH~110 was first
outlined by  \citet{Rei96}. These authors proposed that HH~110 originates from
the deflection (deflection angle $\simeq 60^\circ$) 
of the adjacent HH~270 jet through a grazing collision with a dense
molecular clump of gas. The feasibility of this scenario has been further
reinforced from the results of numerical simulations, which  model the emission
arising from the collision of a jet with a dense molecular clump (see 
\citealp{Rag02} and references therein). In addition, analysis of further data 
are also in agreement with this scenario.  First, a high-density clump of gas
around the region where the HH~270 jet changes its direction to emerge as
HH~110 has been detected through high-density tracer molecules (in HCO$^+$, by
\citealp{Cho01}, and in NH$_3$, by \citealp{Sep10}). Second, results from proper
motion determination are also consistent with the jet/cloud collision scenario
(\citealp{Rei96}; \citealp{Lop05}). A recent work of \citet{Har09} that included
laboratory experiments has given further support to the jet/cloud collision
scenario. 

HH~110 is a good candidate to search for the observational footprints of gas
entrainment and turbulence by analysing the kinematics and the excitation
conditions along and across the jet flow. Some works were performed in the
recent past from long-slit spectroscopy and Fabry-Perot data. The kinematics and
physical conditions, both along the outflow axis and at four positions across
the jet beam, were explored from long-slit spectroscopy by \citet{Rie03a}, who
found very complex structures. \citet{Rie03b} explored the spatial distribution
and the characteristic knot sizes, as well as the spatial behaviour of the
velocity and line width, by performing a wavelet analysis of Fabry-Perot data,
but only covering the H$\alpha$ line. Their results indicated that most of the
H$\alpha$ kinematics can be explained by assuming an axially peaked mean flow
velocity, on which are superposed low-amplitude turbulent velocities. In
addition, their results are suggestive of the presence on an outer envelope that
appears to be a turbulent boundary layer. Finally, \citet{Har09} compared images
from laboratory jet experiments with numerical simulations and with long-slit,
high-resolution optical spectra obtained along HH~110. They found a good
agreement between the shock structures observed in HH~110 and those derived from
experiments of a supersonic jet deflected by a dense obstacle. 

In order to further advance on our understanding of HH~110, this object was included as a target within
a program of Integral Field Spectroscopy of Herbig--Haro objects using the 
Potsdam Multi-Aperture Spectrophotometer (PMAS) in the wide-field IFU mode PMAS
fibre PAcK (PPAK). The data obtained from this IFS HH observing program give
a full spatial coverage of the HH~110 emission in several lines (H$\alpha$,
\nii\ and \sii), thus allowing us to perform a more complete analysis of the
kinematics and physical conditions through the whole flow than all the previous
works. The main results are given in this work. The paper is organized as
follows. The observations and data  reduction are described in \S\ 2.  Results
are given in \S\ 3: the analysis of the physical conditions in \S\ 3.1 and the
analysis of the kinematics  in \S 3.2. A summary with the main
conclusions is given  \S\ 4.

\begin{figure}
\centering
\includegraphics[width=\hsize]{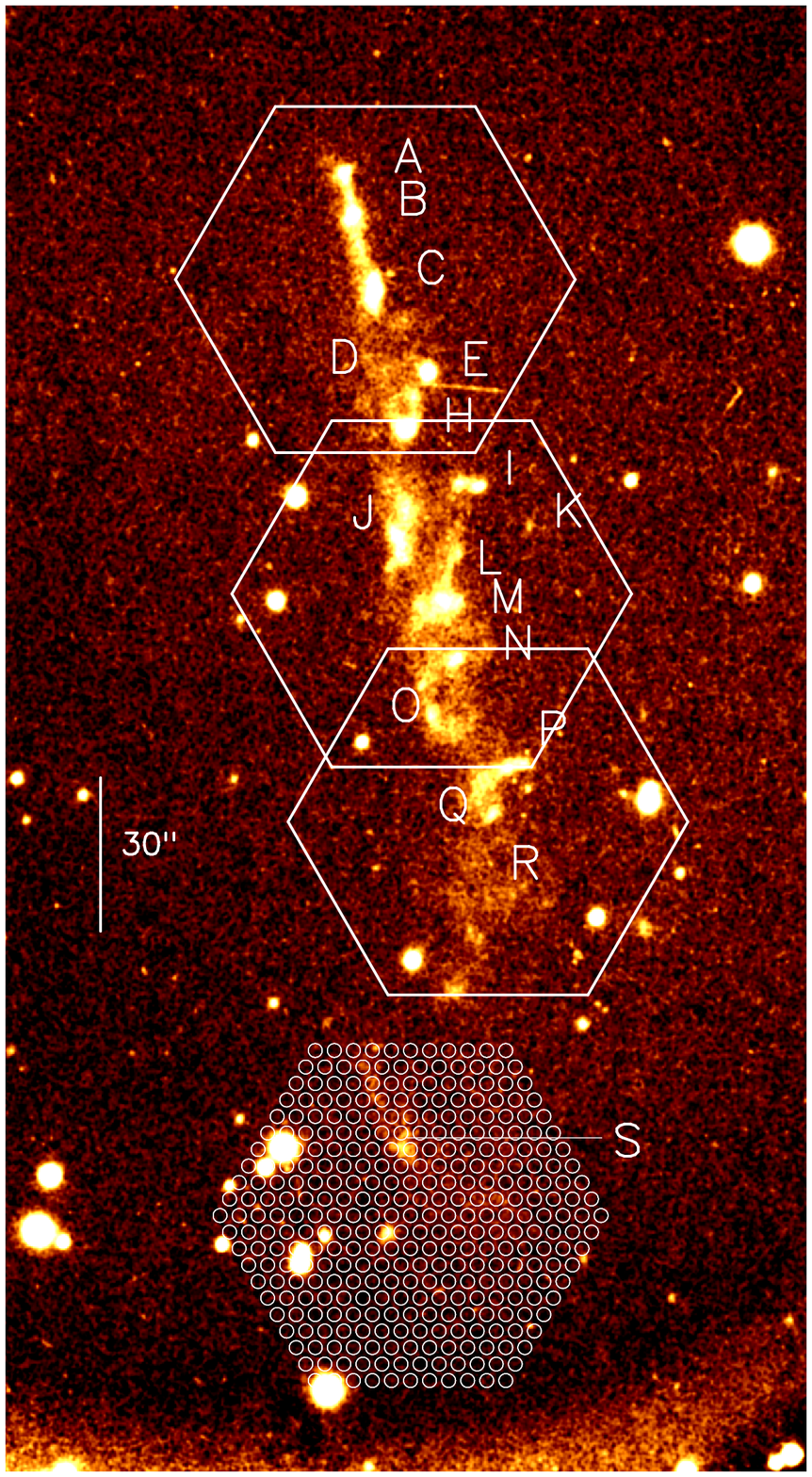}
\caption{CCD image of HH~110 obtained at the Nordic Optical Telescope (NOT)
through a narrow-band filter that includes the \sii\ 6716, 6731 \AA\ lines
(see \citealp{Lop05} for details). The
four IFU pointing fields are superposed on the image. The knots discussed
throughout this work are labeled on the image, according to the nomenclature
first established by \citealp{Rei91}). The contours of the individual fibres
have been drawn on the southern IFU field.
\label{pointings}}
\end{figure}

\section[]{Observations and Data Reduction}

Observations of HH~110 were made on 22 November 2004 with the 3.5-m  telescope of the Calar
Alto Observatory (CAHA). Data were acquired with the  Integral Field Instrument Potsdam
Multi-Aperture Spectrophotometer PMAS \citep{Rot05} using the PPAK configuration that has 331
science fibres, covering an hexagonal FOV of 74$\times$65 arcsec$^2$ with a spatial sampling
of 2.7 arcsec per fibre, and 36 additional fibres to sample the sky (see Fig.\ 5 in
\citealt{Ke06}). The I1200 grating was used, giving an effective sampling  of 0.3 \AA\
pix$^{-1}$  ($\sim15$  km~s$^{-1}$ for H$\alpha$) and covering the wavelength range
$\sim6500$--7000 \AA, thus including characteristic HH emission lines in this wavelength range
(H$\alpha$, \nii\ $\lambda\lambda$6548, 6584 \AA\  and \sii\  $\lambda\lambda$6716, 6731 \AA).
The spectral resolution (\ie\ instrumental profile) is $\sim2$ \AA\  FWHM ($\sim90$
km~s$^{-1}$) and the accuracy in the determination of the position of the line centroid is
$\sim0.2$ \AA\  ($\sim10$ km~s$^{-1}$ for the  strong observed emission lines). 
Four
overlapped pointings were observed to obtain a mosaic of $\sim$ 5\arcmin $\times$ 1$\farcm$5 
to cover  the entire emission from HH~110 (see Fig.\ \ref{pointings}). Table 1  lists the
centre positions of each  pointing, the exposure time and  the HH~110 knots included in each
pointing, following the nomenclature of \citet{Rei96}.

Data reduction was performed using a preliminary version of the {\small R3D}
software \citep{Sa06}, in combination with {\small IRAF}%
\footnote{{\small
IRAF} is distributed by the National Optical Astronomy Observatories, which are
operated by the Association of Universities for Research in Astronomy, Inc.,
under cooperative agreement with the National Science Foundation.} and the
{\small Euro3D} packages \citep{Sa04}. The reduction consists of the standard
steps for fibre-based integral field spectroscopy. A master bias frame was
created by averaging all the bias frames observed during the night and
subtracted from the science frames. The location of the spectra on the CCD was
determined using a continuum-illuminated exposure taken before the science
exposures. Each spectrum was extracted from the science frames by co-adding the
flux  within an aperture of 5 pixels along the cross-dispersion axis for each
pixel in the dispersion axis, and stored in a row-stacked-spectrum (RSS) file
\citep{Sa04}. 
The wavelength calibration was performed  by using the sky
emission lines found in the observed wavelength range. The  accuracy achieved
for the wavelength calibration was better than  $\sim0.1$  \AA\ ($\sim5$
km~s$^{-1}$). 
Furthermore, the large scale diffuse emission from this Orion region was
subtracted from our data by using the signal acquired through the 36 additional
sky fibres.
Observations of a standard star were used to perform a relative
flux calibration. 
 The four IFU pointings were merged into a mosaic using our own routines,
developed for this task (see \eg\ \citealp{San07} and references therein). 
The procedure is based on the comparison and scaling of the relative intensity
in a certain wavelength range for spatially coincident spectra. The pointing
precision is better than 0.2 arcsec, according to the pointing accuracy of the
telescope using the Guiding  System of PMAS in relative offset mode (used for 
these observations). The overlap between adjacent pointings is $\sim$
10--15\%  the field of view (more than 30 individual spectra). 
The merging
process was checked to have little effect on the  accuracy of the
wavelength calibration of the final datacube making use of the sky lines (\ie\ a
common reference system) present in the spectra.
A final datacube  containing the 2D spatial plus the spectral
information of HH~110 was then created from the 3D data by using {\small Euro3D}
tasks  to interpolate the data spatially until reaching a final grid of 2
arcsec of spatial sampling and with a spectral sampling of 0.3 \AA\ . Further
manipulation of this datacube, devoted to obtain  integrated emission line maps,
channel maps and position--velocity maps, were made using several common-users
tasks of {\small STARLINK}, {\small IRAF} and {\small GILDAS} astronomical
packages and {\small IDA} \citep{Ga02}, a specific {\small IDL} software to
analyse 3D data.

\begin{figure*}
\centering
\includegraphics[angle=-90,width=0.95\hsize]{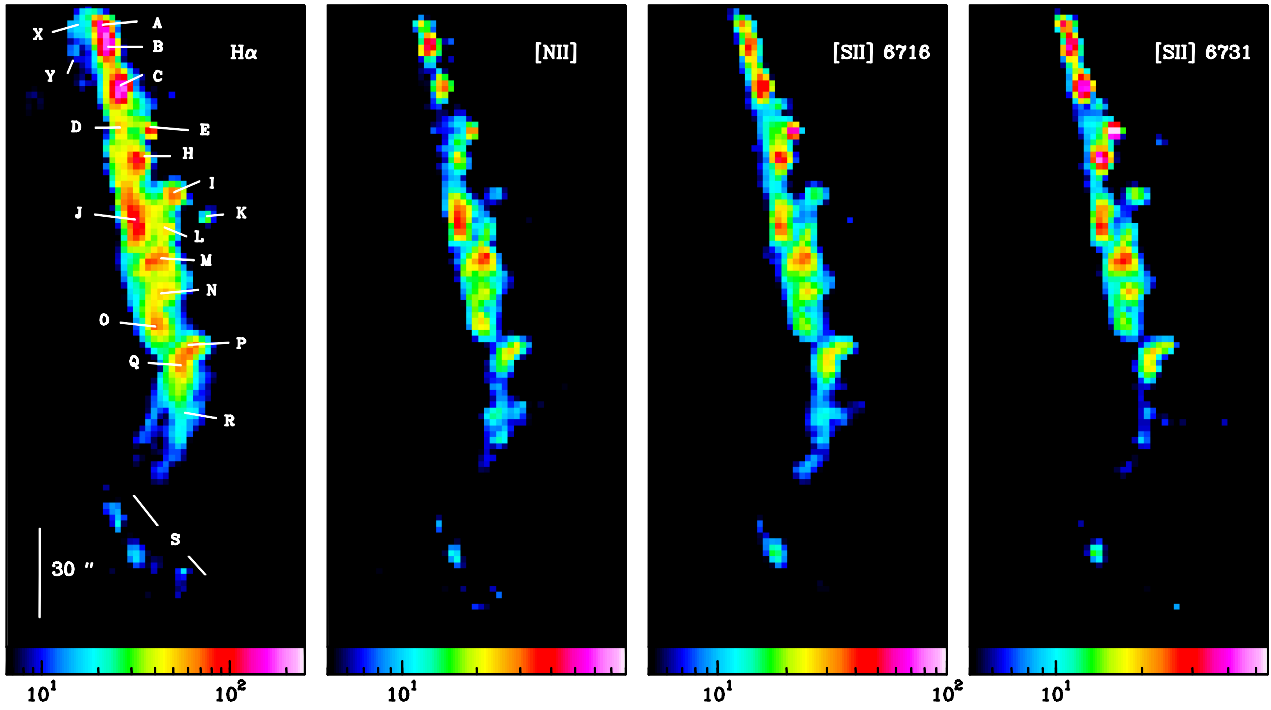}
\caption{
Integrated maps of the HH~110 emission obtained from the datacube 
by
integrating the signal within the wavelength range including the 
line labeled in each panel:
(H$\alpha$ : $\lambda$6558.3--6565.9 \AA, 
\nii\ :	    $\lambda$6579.6--6585.8 \AA,  
\sii\  $\lambda$6716:  6712.8--6718.1 \AA\  and 
\sii\  $\lambda$6731:  6727.2--6733.1 \AA ).  
Fluxes have been displayed in a logarithmic scale, in  units of 10$^{-16}$
erg~s$^{-1}$~cm$^{-2}$. In all the
maps, north is up and East is to the left. 
The HH~110 knots and the spatial scale have been labeled in
the H$\alpha$ panel (30 arcsec $\simeq$ 0.07 pc for a distance of 460 pc; \citealp{Rei91}).
\label{flux}}
\end{figure*}

\begin{table}
\caption{Positions of the single pointings of the HH~110 mosaic.
\label {tjcmt}}
\begin{tabular}{@{}lccc@{}}
\hline
\multicolumn{2}{c}{Position}&Exp. time(s)&knot \\
\hline
05$^{\rm h}$51$^{\rm m}$25$\fs3$&+02$\degr$55$\arcmin$16$\arcsec$&1800&A-E \\
05$^{\rm h}$51$^{\rm m}$24$\fs4$&+02$\degr$54$\arcmin$20$\arcsec$&1800&I-N  \\
05$^{\rm h}$51$^{\rm m}$23$\fs8$&+02$\degr$53$\arcmin$41$\arcsec$&2400&N-R  \\
05$^{\rm h}$51$^{\rm m}$24$\fs9$&+02$\degr$52$\arcmin$40$\arcsec$&2700&S    \\
\hline
\end{tabular}
\end{table}

\section{Results} 

\subsection{Physical conditions}

\subsubsection{Morphology}

Some morphological differences have been found in previous narrow-band 
images of HH~110 between the
H$\alpha$+\nii\ and \sii\ emissions. In an attempt to explore these 
differences, we  obtained
IFS-derived narrow-band images for each of the
emission lines included in the observed spectral range (H$\alpha$, \nii\
$\lambda$ 6584 \AA\ and \sii\ $\lambda$ 6716,6731 \AA). For each position, the
flux of the line was obtained by integrating the signal over the
wavelength range of the line ($\lambda$ 6558.3--6565.9 \AA\ for H$\alpha$; 
$\lambda$ 6579.6--6585.8 \AA\ for \nii; $\lambda$ 6712.8--6718.1 \AA\ for \sii\
$\lambda$6716 and  $\lambda$ 6727.2--6733.1 \AA\ for \sii\ $\lambda$6731),  and
subtracting a continuum, obtained from the adjacent wavelength range free of
line emission. The narrow-band maps of HH~110 obtained from the IFS data 
(Fig.\ \ref{flux}) are in good agreement with
those obtained for this jet through  ground-based, narrow-band
images by different authors (see \eg\ \citealp{Rei96};
\citealp{Lop05}). However, since the IFS data allow us to properly isolate the
H$\alpha$ line emission  from  the \nii\ lines, we obtained the
first known image  of HH~110 in the  \nii\ emission. In this sense, the IFS maps are
better suited for comparing the jet morphology in the H$\alpha$ and \sii\
emissions, since the H$\alpha$
emission is not affected by the contamination from \nii\ lines.

Most of the HH~110 knots were detected in all the emission lines,  although the
brightness varies among different lines, being a signature of the excitation
conditions, as we will discuss later. There are a few emission features whose
brightness  in the \nii\ and \sii\ lines should be very weak relative to the
H$\alpha$ brightness and  were barely detected or not detected at all in the \nii\
and \sii\ lines,  namely: (i) the knot K, appearing quite isolated towards the
western outflow edge, but outside the lower-brightness emission surrounding the
knots, and (ii) the extended features towards the east of knots A to C, labeled X and
Y. These faint H$\alpha$ features were first detected in the  
ground-based,
narrow-band image of \cite{Rei91}.  Our new IFS data indicate the X and Y features to be a
''true''  emission in the H$\alpha$ line, having neither significant contributions 
from the extended  nearby continuum (that was removed) nor from \nii\ lines 
(if any,
it should be below the 3$\sigma$ rms level). 

The knots are found surrounded by a lower-brightness nebular emission for all the emission
lines. This more diffuse emission also shows a shock-excited spectrum, and likely arises
from the gas in the
boundary layer that  is being  incorporated into the outflow.  The large scale, low-velocity
diffuse emission component detected in the Orion region was removed from our data and is not
contributing to this low-brightness, interknot emission structure we are referring to.
Figure\ \ref{flux} shows that the cross section along all of the jet appears wider
in H$\alpha$ than in the other lines, which all have similar spatial widths. 
Since with the present spatial resolution the sizes of the knots are similar 
at all lines, the wider H$\alpha$ cross-section of the jet 
should arise from the contribution of the low-brightness, nebular  emission.  
In our maps, this weaker emission  appears  spatially
more extended in H$\alpha$ than in the rest of the  observed lines. This fact could be due, in
part, to the weakness of the \nii\ and \sii\ emissions, which makes the emission undetectable
in our maps (up to the 3$\sigma$ rms level) farther away from the knots.

\begin{figure*}
\centering
\includegraphics[angle=-90,width=0.75\hsize]{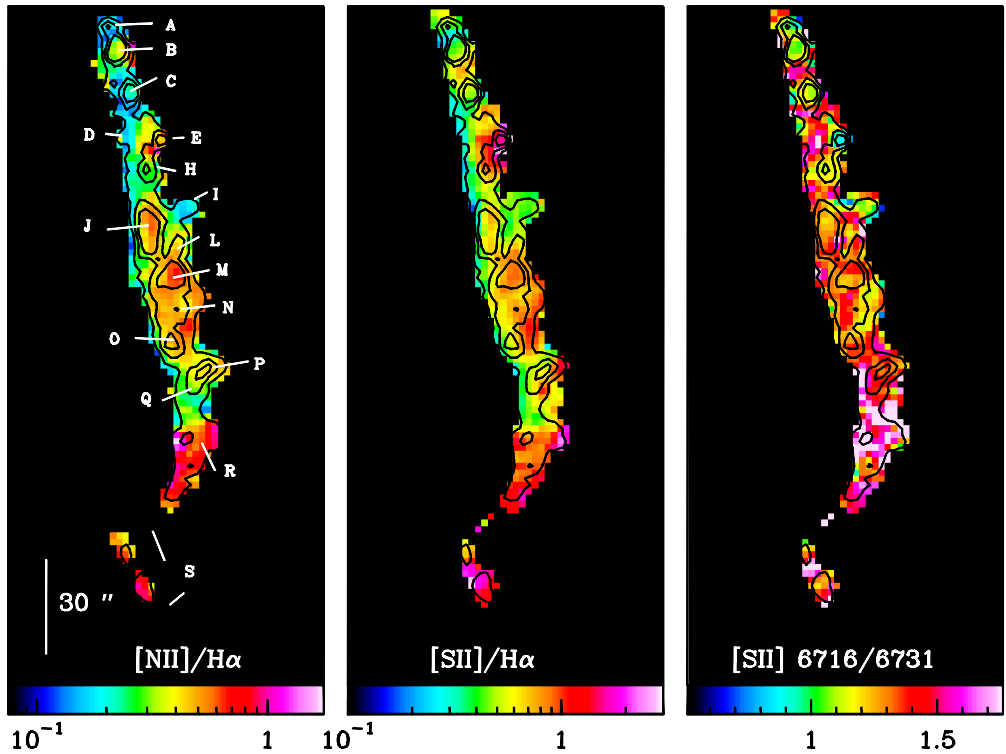}
\caption{Line-ratio maps derived from the integrated line emissions displayed in
Fig.\ \ref{flux}. The \sii/H$\alpha$ and \nii/H$\alpha$ maps show the excitation
conditions of the HH~110 emission, while the \sii\ 6716/6731 map traces the 
electron density. Contours of the integrated flux of the H$\alpha$ line have
been overlaid to
help with the knot identification, which have been labeled in the \nii/H$\alpha$ map.
\label{cfi}}
\end{figure*}

\begin{figure*}
\centering
\includegraphics[angle=-90,width=0.75\hsize]{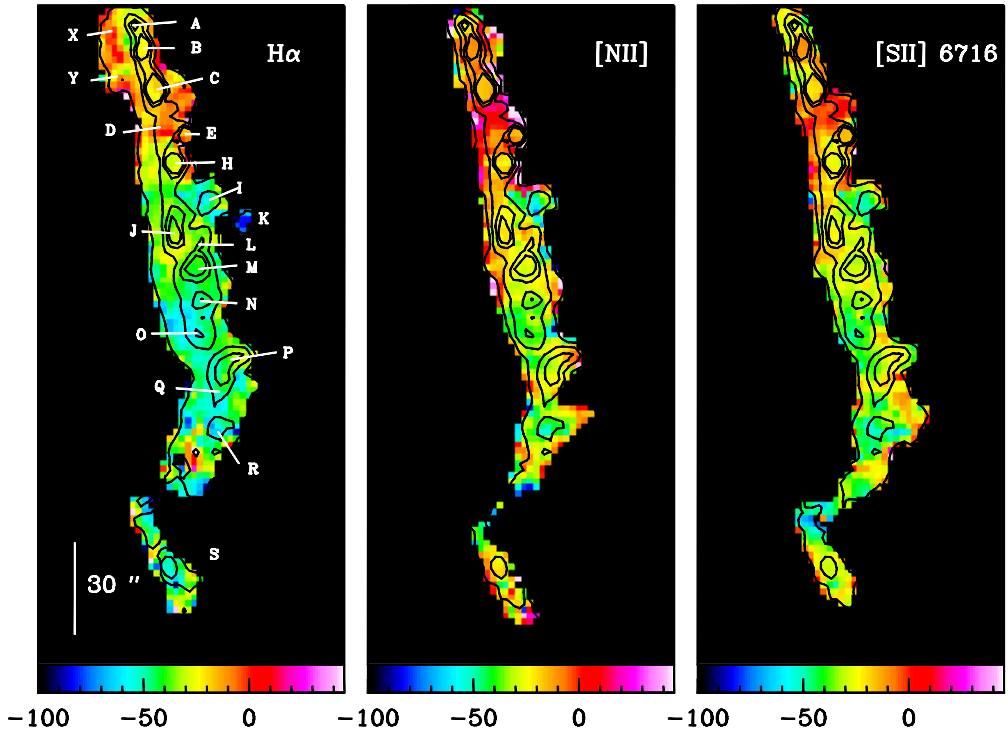}
\caption{
 Flux-weighted mean velocity maps for the line labeled in the corresponding 
 panel. 
Velocities (in km~s$^{-1}$) are referred to the
\vlsr\ of the parent cloud. Contours and knot identification, as 
in Fig.\ \ref{cfi}
\label{vel}}
\end{figure*}

\subsubsection{Excitation and density conditions}

First results on the physical conditions of HH~110 were obtained through
 long-slit spectroscopy  acquired along the jet outflow 
(\citealp{Rei96}; \citealp{Rie03a}) and at a several positions across the jet
beam (\citealp{Rie03a}). 
 The widths of the slits used in these observations were 2 and 1.5 arcsec, 
respectively.
In both works, the \sii/H$\alpha$ ratios derived for
the knots intersected by the slit along the jet axis outflow  range from
$\sim$0.2 to $\sim$0.7. These ratios correspond to an intermediate/high
degree of excitation (\citealp{Rag96}). 
A trend was found of  an
increasing   \sii/H$\alpha$ ratio (hence a decrease of the excitation) moving 
from knot C to knot M, where the lowest excitation was detected. A decrease
of this ratio (hence an increase of the excitation) was found beyond knot M.
In general, the
values derived in both works are  well consistent, except for knots A and N,
where the \sii/H$\alpha$ ratio reported by \citet{Rie03a} is significantly
higher (by a factor of two) than the ratio reported by \citet{Rei96}. This is
not surprising since the excitation conditions should change significantly within arcsecs
across the jet beam, as first revealed by the
spectroscopy of the jet cross section performed  by \citet{Rie03a}. Thus, the
sampled regions of knots A and N might not coincide in both
observations.  Furthermore, no clear trends for the excitation structure of the
jet cross sections were found from the spectra acquired by \citet{Rie03a}. These authors 
found a complex structure at the four positions across the jet beam, 
where the spectra were obtained.

To look for a more complete picture of the entire spatial structure of the
excitation and density of the jet, we obtained   the \sii(6716+6731)/H$\alpha$
and  \nii(6548+6584)/H$\alpha$  line-ratio maps from the IFU data (Fig.\
\ref{cfi}). The line-ratio maps were created by dividing  spaxel-by-spaxel the
corresponding flux-integrated line maps (all the spaxels with  values below
5$\sigma$  were blanked before to obtain the line-ratio map). 
At the weakest  knots, the estimated
uncertainties  are 10\% and 15\% in the  \sii(6716+6731)/H$\alpha$ and
\nii(6548+6584)/H$\alpha$ line-ratio values respectively.

The \sii/H$\alpha$ and  \nii/H$\alpha$ maps clearly reveal the complex
excitation structure of the emission that was only barely outlined from
long-slit spectroscopy. Several facts should be remarked: (i) for  the knots
that were sampled through  long-slit spectroscopy, the line-ratio values derived
from the IFS data around the peak positions are in good agreement with those
derived  from the long-slit spectra. For knots A and N, the ratios obtained from
the IFS data are in better agreement with those derived  by \citet{Rie03a} than
with those derived by \citet{Rei96}; (ii)  IFS  line-ratio maps revealed some
regions in which  the degree of gas excitation is low.  In particular, the
highest \sii/H$\alpha$ ratios ($\ge$ 1.5) are found around knot H and slightly
lower ratios, which are also compatible with a low degree of excitation, are
found around knots R and S. This wider range of excitation degree could  not be
detected by previous long-slit observations, which did not sample  these low
excitation regions; (iii) no  clear systematic trend in the spatial distribution
of the excitation was found neither along the jet outflow nor across the jet
beam. However, the integrated \sii/H$\alpha$ and \nii/H$\alpha$ maps indicate 
that the excitation is higher for the northern jet knots A to D (where the
outflow has a higher  collimation) than beyond knot E, from  the region where
the outflow widens and the knots begin to lose their alignment relative to the
axis direction traced by knots A-C. For this jet region (\ie\ from knots E to Q)
the line ratio maps show some signatures of an increment in the  excitation
moving outwards, from the knot peak to the eastern knot border and to the
interknot. The line ratio maps are thus suggestive of a decrease in the
excitation from east to west across the jet beam. This is true for all knots,
except around knot I (at the western jet border), where the excitation is higher
and similar to that derived in the northern (A to D) knots.

The spatial distribution of the electron density ($n_\rmn{e}$)  is traced by
the \sii\ 6716/6731 map.  Figure \ref{cfi} shows the map  created from the IFU
data (with the assumptions mentioned before for the excitation maps). At the
weakest knots, the estimated uncertainty is 15\%.
The values found range from $\sim$
0.9 to $\sim$ 1.4, which correspond  to $n_\rmn{e}$ ranging from 1000 to 50
cm$^{-3}$ respectively (these values were derived using the {\small TEMDEN} task
of the {\small IRAF/STSDAS} package, and assuming $T_\rmn{e}=10^4$~K). The
electron density cannot be properly evaluated around the knot R region, because
the derived line ratios are higher than the values for which the \sii\ 6716/6731
is density sensitive in the low-density limit. 
In general, the trend found 
is a decrease in the  $n_\rmn{e}$, moving along the jet from north to south. 
This result is consistent  with
the density behaviour of the knots derived from the long-slit spectra  by  \citet{Rei96} and 
\citet{Rie03a}.  
However, the spatial distribution of the density obtained through IFS (Fig.\
\ref{cfi}) is more complex than that
outlined from long-slit spectroscopy, and could not be properly described based
on this kind of observations.
The density map indicates that, in general, the
electron densities are higher around the knot peak positions and decrease
towards the edges of the knots  and towards the interknot gas.  Across the jet
beam, a  trend is observed in the knots peaking towards  the west side of the
axis defined by knots A--C  (\eg\ knots E, I, N, P) that are denser than
knots peaking  towards the east side of this axis (\eg\ knots D, J, O, Q).  
The density map also shows some local departures of this general trend that are
significant. In particular, it should be noted that the highest $n_\rmn{e}$ is
reached around knot E (\ie\ $\sim$~40~arcsec south from knot A).  In fact, the
IFU data revealed an enhancement in the electron density, reaching $n_\rmn{e}$
values of $\simeq$~1000~cm$^{-3}$ and with a maximum of  $\simeq$~1300 cm$^{-3}$
located $\simeq$ 3 arcsec southwest of the knot E peak intensity. This denser
clump, extending over $\simeq$ 30 arcsec$^2$,  appears well delimited at its
eastern edge, as shown by a sharp fall of $n_\rmn{e}$, which reaches values of
$\simeq$ 200--300~cm$^{-3}$ by only moving $\simeq$ 2 arcsec east from the E
peak intensity.  In addition,  although $n_\rmn{e}$ gently decreases from knots
A ($\sim$~600~cm$^{-3}$) to H ($\sim$~400~cm$^{-3}$), the knot I, southwest of
all of these knots, has a $n_\rmn{e}$ similar to knot B ($\sim$~500~cm$^{-3}$),
while knot D, southeast of knot C, has  similar low values
($\sim$~150~cm$^{-3}$) to those found at the southern, more rarefied  knots, 
beyond knot N.

\subsection{Kinematic analysis}

\begin{figure}
\centering
\includegraphics[angle=-90,width=\hsize]{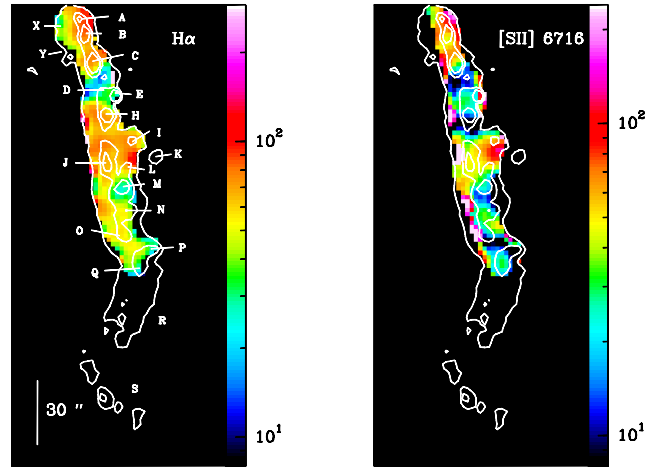}
\caption{Maps of the FWHM (in km~s$^{-1}$) for the line labeled in the corresponding panel.
Contours and knot identification, as in Fig.\ \ref{cfi}. Spaxels with a FHWM
below 115~\kms have been blanked. (At several positions,
like in the southern knots,
FWHM should be narrower and could not be properly determined due to the
instrumental line profile).
\label{fwc}}
\end{figure}

\subsubsection{Spatial velocity distribution}

The emission lines at most of the positions mapped in HH~110 show asymmetric
profiles that cannot be properly fitted by a single Gaussian. Hence, we
calculated the flux-weighted mean radial velocity (or first-order intensity
momentum) instead of the line centroids of  a Gaussian fit,
and the flux-weighted rms width of the line (or second-order intensity
momentum), proportional to the FWHM of the line, and 
described
the integrated velocity field of the jet\footnote{All the
velocities in the paper are referred to the local standard of rest (LSR)
velocity reference frame.  A  $\mbox{\vlsr}=+8.5$~\kms\ for the parent cloud has
been taken from \citet{Rei91}.} from these intensity momenta. 
HH~110 shows a velocity field behaviour very similar in all the mapped
lines (Fig.\ \ref{vel}). 
The radial velocities  are blueshifted 
($\mbox{\vlsr}$ ranging from $\sim$~--10 to --80~\kms\ in H$\alpha$), with an increase towards
more blueshifted values moving from north to south. 
 Velocities that were derived from H$\alpha$ are 
 blueshifted by $\sim$~20~\kms relative to those derived from the \nii\ 
 and \sii\ (the velocities
derived from these two lines are  similar). 
 The velocity offset found between H$\alpha$ and
\sii\ is not an artifact caused by an inaccurate wavelength
calibration of the IFS data, since it has been also found in long-slit, higher
resolution data  (\eg\ \citealp{Lop05}). 
A more detailed inspection of the velocity maps reveals a more
complex structure of the velocity field: (i) the less blueshifted velocities
($\mbox{\vlsr}$ $\sim$~--10 to --15~\kms\ in H$\alpha$) are found around the
northern emissions X, Y and in knot D, all of them located towards the eastern
side of the jet axis direction traced by the knots A--C; (ii) conversely, more
blueshifted velocities ($\mbox{\vlsr}$~$\sim$~--50 to --60~\kms\ in H$\alpha$)
appear around the regions of knot O and Q, which also are  located at the
eastern jet side. Even more  blueshifted velocities (up to
$\mbox{\vlsr}$~$\sim$~--80~\kms\ in H$\alpha$) are found around knots I and K,
located at the western jet border, quite outside of the main jet body.

The maps of the FWHM for the H$\alpha$ and \sii~$\lambda$~6716 \AA\ lines 
obtained from the second-order intensity momentum are shown in Fig.\ \ref{fwc}.
 Maps were corrected from the instrumental profile by subtracting it in
quadrature spaxel-by-spaxel. Spaxels with values below a 5$\sigma$ rms level  
were blanked.
The spatial distribution of the FWHM
shows a complex pattern in both lines. 
In general, widths are broader for H$\alpha$ than for the
\sii\ lines. The FWHM of the H$\alpha$ ranges from 30 to 90~\kms, being 
in good agreement 
with those found by \citet{Rie03b} from  Fabry-Perot data
and with those derived by \citet{Har09} from long-slit, high-resolution spectra. 
The northern knots (A to C) present the largest  values (80--90~\kms)  of the
FWHM, decreasing towards the south of the outflow (up to 30--40~\kms\
around knots M and Q). In fact, the FWHM around the arc-shaped, southern S
emission region should be even narrower, but the widths  could not be properly 
evaluated due to the instrumental line profile. Like for other jet properties, 
this general trend is not well fulfilled by the whole outflow: a region having
significant narrow lines, their widths being close to detection limit,  is also 
found  around knot D, while higher  FWHM values (90~\kms,  similar to the widths
derived around knot A) are found around knot I. Concerning the FWHM of the \sii\
lines, their values  range from 20 to 70~\kms and are in  agreement with
the results of \citealp{Har09}). The spatial distribution of the FWHM values for
the \sii\ lines are similar to the FWHM H$\alpha$ map.  Some key differences
in FWHM between H$\alpha$ and \sii\ 
should be mentioned: (i) the widths of the \sii\ lines drop significantly around
knot C, where its value diminished by a factor of two relative to that found
around knot A,  while knots C and A have  comparable FWHMs in H$\alpha$; (ii)
knots M and Q have very similar FWHM values in \sii\ and H$\alpha$, while the
general trend for the rest of the knots is the H$\alpha$ line being wider (by
15--20~\kms typically) than the \sii\ line.

\subsubsection{Peak intensity knot velocities}

\begin{figure}
\centering
\includegraphics[width=1.01\hsize]{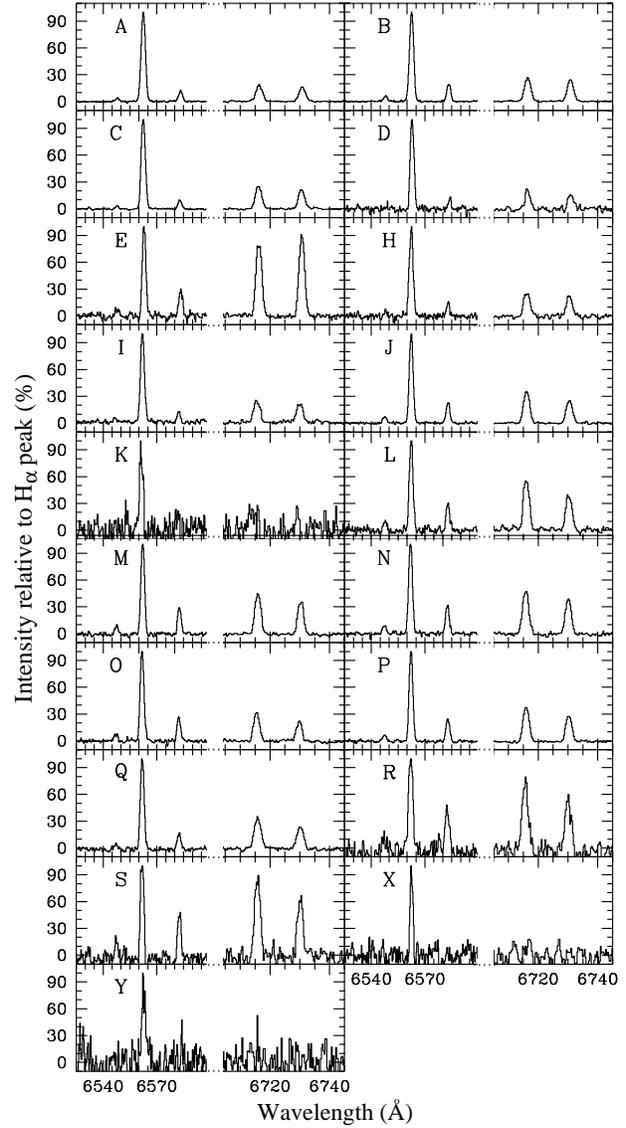}
\caption{Spectra of the HH~110 knots in the spectral windows including H$\alpha$
+ \nii\ and \sii\ 6716, 6731 lines. Each spectrum has been extracted for the
spaxel that
corresponds to the peak
intensity position (\ie within an aperture of 2~arcsec) 
and its intensity has been normalized to the H$\alpha$ peak
intensity.
\label{espknotp}}
\end{figure}

In an attempt to compare long-slit and IFS data, we have obtained   a
representative spectrum for each jet knot by extracting the spectrum of the
spaxel (aperture of 2~arcsec) that corresponds to the position of the knot
intensity peak (Fig.\ \ref{espknotp}).  The knots show significant differences
in their physical conditions, as inferred from their relative line intensities.
Moreover, the lines present asymmetric profiles that in some knots are
suggestive of being double-peaked (see, \eg\ knots H, I) although the profiles
cannot be resolved because of the spectral resolution of the data.

\begin{figure}
\centering
\includegraphics[width=1.02\hsize]{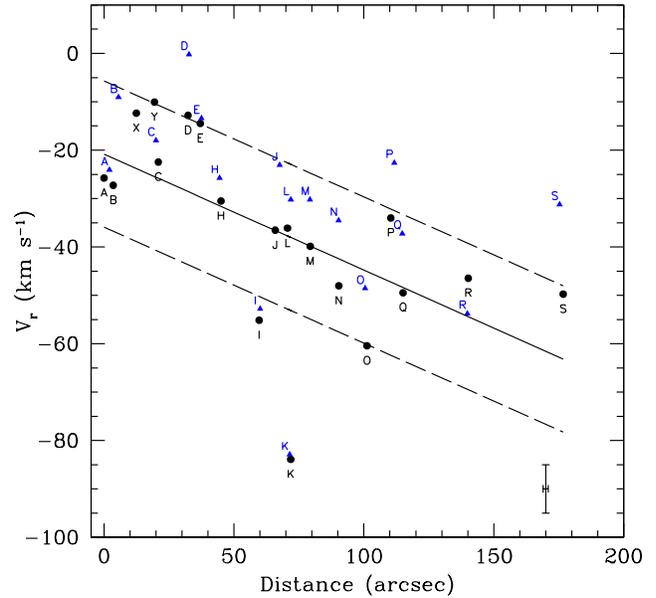}
\caption{
Radial velocities measured at the position of the peak intensity of 
the HH~110 knots, as a function of distance to the knot A peak intensity in 
H$\alpha$. The black dots correspond to the values derived from H$\alpha$ and 
the blue triangles, from the \sii\ lines. The typical value of the velocity
error is $\sim$ 10~\kms and the uncertainty in the knot position $\sim$ 1
arcsec, and are displayed in the lower-right corner of the figure. 
The straight lines mark the linear fit to the data (solid line) and the 1~$\sigma$
deviation from the fit (dashed lines).
\label{velpeak}}
\end{figure}

Figure \ref{velpeak} displays  the radial velocities at the position of the knot peak
intensity, in both H$\alpha$ and \sii\ lines, as a function  of the distance to knot
A.  The two radial velocity fields are very similar.  A trend of increasing
blueshifted  (more negative) velocity values with the distance to knot A  is
outlined. This trend is consistent with the velocity behaviour found in previous
works (\citealp{Rie03b}; \citealp{Har09}).  The radial velocities of some knots
present deviations of this general trend. 
In particular, the isolated knot K (found outside the jet body) has a
blueshifted radial velocity more negative than the rest of the knots, and its
deviation  from the general trend is statistically significant. A linear fit of
the peak radial velocity of the H$\alpha$ line with distance,	including all the
knots (19), was obtained (correlation coefficient $r= 0.61$; rms fit residual
$\sigma=15.1$ km s$^{-1}$). The fit is shown in  Fig.\ \ref{velpeak} by a continuous
straight line, and the $1\sigma$ deviation from the fit, by the dashed lines.
Knot K deviates from the fit by $3\sigma$. This deviation is statistically
significant for a sample of 19 knots. Then, a new linear fit was calculated
after removing knot K from the sample (i.e.\ 18 knots). We obtained values of
$r=0.74$ and $\sigma=10.5$ km s$^{-1}$. The knots that deviate more than
$1.5\sigma$ from this new fit are knot O ($1.8\sigma$ deviation), I ($2.2\sigma$
deviation), and K shows now a $4.6\sigma$ deviation. Unfortunately there is no
proper motion determination for knot K that would allow us to calculate the
projection angle of its motion. Hence, the current data do not allow us to
discern whether the kinematic behaviour of knot K, with a radial velocity
appreciably blueshifted as compared with the rest of the knots, can be only
attributed to the jet geometry (i.e.\ to projection effects). Recently, \citet{Har09}
imaged the HH 110 region in the H$_2$ 2.12 $\mu$m line. They
report a weak jet emerging from the source IRS1 (to the west of HH 110) and
extending from north to south. This weak jet seems to be pointing at knot K. In
conclusion, there are some indications (a location isolated from the jet flow,
and a kinematics significantly different from the rest of the knots) that knot K
could be related to IRS1, but its relationship with the HH 110 outflow cannot be
fully discarded based on the results presented in this work.

\begin{figure}
\centering
\includegraphics[width=1.01\hsize,clip]{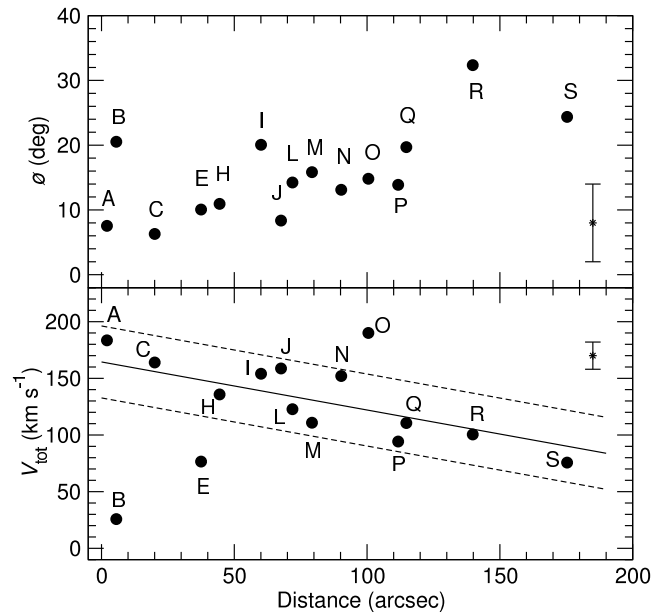}
\caption{Total velocities (lower panel) and inclination angle, defined by the
direction of the knot motion and the plane of the sky (upper panel) as a
function of distance to the position of the \sii\ intensity peak of knot A.
The typical values of  $\phi$ and \vtot\ errors are $\sim$~6$^\circ$ and 
$\sim$~12~\kms, respectively and are displayed at the corner of each panel. 
The uncertainty in the knot position ($\sim$ 1 arcsec) 
is smaller than the size of the
symbols.
Straight lines that mark the linear fit to the data (solid line) and the 1~$\sigma$
deviation from the fit (dashed lines) are drawn on the velocity panel.
\label{angulo}}
\end{figure}

The full spatial velocity (\vtot) and an upper limit for  the angle $\phi$ between
the knot motion and the plane of sky ($\phi$=0$^\circ$ when the flow moves in the
plane of the sky) were calculated from the knot  proper motions, derived  from
multiepoch \sii\ narrow-band images (\citealp{Lop05}) and the radial velocities,
derived in this work from the \sii\ lines at the position of the  knot  intensity peak
(the typical errors for $\phi$ and \vtot\  are $\sim$~6$^\circ$ and 
$\sim$~12~\kms\ respectively). The results are displayed in Fig.\ \ref{angulo}. 

The values derived for the inclination angle suggest that the flow remains close to the
plane of the sky ($\phi~\leq$~20$^\circ$ for most of the knots).  The inclination  gently
increases from $\phi~\simeq$~8$^\circ$ at the beginning of the outflow (around knot A) to 
higher values ($\geq$~30$^\circ$) for distances $\geq$~150~arcsec from knot A (around knots R
and S, coinciding with the region where the outflow has a highly curved shape).

As can be seen from Fig.\ \ref{angulo} (lower panel),  the spatial velocity (\vtot) derived at
most  knots is  $\geq$~100~\kms. There is a trend of decreasing velocity with distance  along 
the outflow, from \vtot~$\simeq$~185~\kms\ for knot A  to \vtot~$\simeq$~75~\kms\ for knot S. 
A few knots present a deviation from this velocity trend that might be significant: knot O,
with a higher velocity (\vtot $\simeq$ 190~\kms) than the neighbouring knots (\vtot $\leq$
150~\kms), and knots E and B,  having lower \vtot\ than the neighbouring knots. In
particular,  the lowest spatial velocity is obtained at knot B  (\vtot $\simeq$ 25~\kms). This
value is far from the spatial velocities of its neighbouring knots (\vtot $\geq$ 165~\kms) and
even far from the velocities of the slower knots.   In order to better visualize the velocity
behaviour to we are referring,   a  linear fit was  calculated after removing  knot B from the
sample (\ie\ 14 knots). The fit (correlation coefficient $r=0.53$;  rms fit residual
$\sigma$=31.8~\kms) has been drawn in Fig.\ \ref{angulo} by a continuous straight line, and
the 1$\sigma$ deviation from the fit, by the  dashed  lines. As can be seen from the figure, 
the spatial velocity deviates  $\leq$ 1$\sigma$ for most of the knots.  A few knots  lie
outside the region delimited by the dashed lines: knot O (2.2$\sigma$ deviation above the
fit), knot E (2.3$\sigma$ below the fit)  and knot B  (4.3$\sigma$ below the fit).  The
region around knot B could correspond to the region where the outflow is excavating a channel
through the dense molecular clump (\citealp{Rei96}; \citealp{Lop05}) and might trace the
location where the jet collides with the cloud.  Hence, significant deviations from the
kinematics trend that account for the jet/cloud interaction should be expected at this
location. This may explain the appreciably low spatial velocity found for knot B. Concerning
the other two knots with deviations less significant, knot O   also has a highly blueshifted
radial velocity, (although within a 1$\sigma$ deviation from the radial velocity fit, see
Fig.\ \ref{velpeak}). The low \vtot\ at knot E mostly corresponds to the tangential velocity
component (see also Fig.\ \ref{velpeak})  As was mentioned before, the highest values of the
electron density in the outflow, measured from the integrated \sii\ line  ratio, are found
around knot E location. Thus, similarly to what happens around knot B, a stronger interaction 
between the jet and the inhomogeneous ambient material around  the knot E location might be
producing these departures in the kinematics and physical conditions relative to those found
at the surrounding knots.

\subsubsection{Line Profile Shapes}

\begin{figure}
\centering
\includegraphics[width=1.05\hsize]{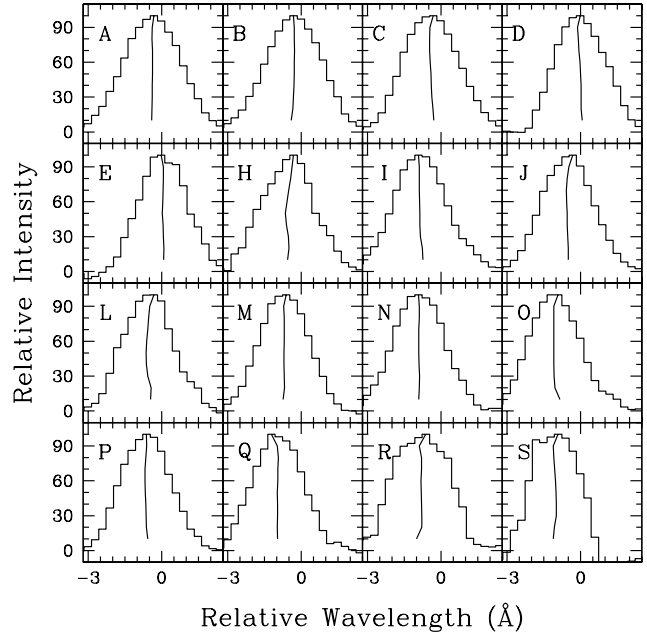}
\caption{H$\alpha$ line profiles of the knots labeled in each panel, obtained from the
spectra extracted at the position of the knot peak intensity (see Fig.\
\ref{espknotp}). 
The line bisector has been drawn for each profile.
In each panel, the line intensity  has been normalized 
to the value at the peak. On the x-axis, the wavelengths are relative to the
rest H$\alpha$ wavelenght.
\label{bise_ha}}
\end{figure}

As already mentioned in \S 3.2.1,  the line profiles show some degree of asymmetry
relative to a single-Gaussian shape that varies from knot to knot.  This can be seen
for the H$\alpha$ and \sii\  lines of the spectra extracted at the position of the
peak intensity of each knot (Fig.\ \ref{espknotp}).   The analysis of spectral
line asymmetries through line bisectors provides a strong tool to understand the
shape and origin of line profiles, and is widely used in stellar astrophysics (see
\eg\ \citealp{Que01}; \citealp{Dal06}; \citealp{Mar08}). In order to make more clear
the asymmetries of the HH~110 line profiles,  we calculated the line bisectors for
the H$\alpha$ line at the position of the knot peak intensity. 
The results  are shown
for each profile in Fig.\ \ref{bise_ha}. We found knots with some high-velocity
emission excess, as shown by the tilt of the line bisector towards blueshifted
wavelengths (knots A, C, D, H, J, L, M, O, P, R and S).   We will refer to these
knots as those having ''blue asymmetry". For other knots, there is  low-velocity
emission excess, with the line bisector tilted  towards redshifted wavelengths
(referred as knots having ''red asymmetry'': knots E, I and Q).  Finally, a few knots 
show a  nearly symmetric profile (knots B and N). Such line-profile behaviour might
be explained in part by the combined effect of the knot velocity and flow inclination
at the knot position. 

Some clues can be obtained from  the work of \citet{Har87}, who modeled the effect
of varying the velocity and the  inclination angle of the flow in the plane of the
sky on the  H$\alpha$ line profiles emerging from bow shocks. They found that the
asymmetry of the line profile increases with the velocity of the shock, and with the
inclination angle.  Resolved, double-peaked profiles are only found  for velocities
higher than $\simeq~150$~\kms and for angles greater than $\simeq~45^\circ$. 
Furthermore,  moderate variations on the inclination angle result in significant
effect in the shape of the line profile. For a given angle, the shape of the
asymmetry changes from ''red'' to ''blue'' depending on the velocity: \eg\ for low
inclination angles ($\leq~30~^\circ$), a ''red asymmetry'' is obtained for low
velocities  ($\leq 100$ km~s$^{-1}$) and a ''blue asymmetry'' for higher velocities 
($\simeq 200$ km~s$^{-1}$).

Interestingly, most of the HH~110 knots showing a ''blue asymmetry'' in the line
profile correspond to knots having high velocity (\vtot\ $\geq$~160~\kms,~\eg: A,
C, J, O) or inclination angle $\phi~\simeq~30^\circ$  (\eg: R, S). ''Red
asymmetry'' is  observed  in  knots  with lower velocities (\eg\  E, with
\vtot~=~80~\kms and Q, with \vtot~=~110~\kms) and lower inclination angles
($\phi~\leq~20^\circ$).  Note in addition that some pairs of knots with close
values of \vtot\ and $\phi$ show very similar profiles, as is the case of knots C
and J, and of knots H and L. 

Hence, there is a general agreement  of the  H$\alpha$ line profiles  observed in HH 110 with
the obtained from the shock models of \citet{Har87}. However, other properties of the emission
could also being contributing to the line shapes. For example, different morphologies  of the
bow emission, keeping all the other parameters constant, can give different line profiles, as
shown by \citet{Sch05}. They found that the bluntness at the apex and the radius of the tail
are two important factors in determining the line profiles for the case of H$_2$ lines arising
from a bow-shock. In fact, \citet{Har09} found from laboratory jet experiments how the working
surface of a deflected bow shock  present complex emission structures as those observed in
HH~110. These structures  give rise to different asymmetric line profiles,  depending on the
HH~110 knot and hence, on the shock velocity and projection angle.

\begin{figure}
\centering
\includegraphics[width=\hsize]{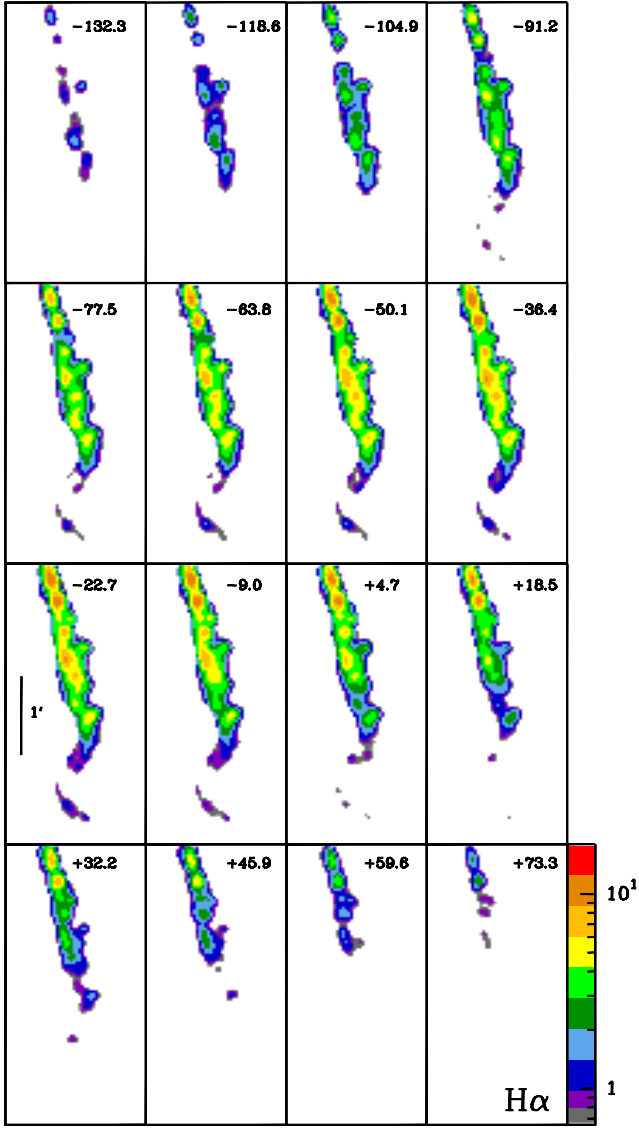}
\caption{
H$\alpha$ channel maps centred on the  \vlsr\  radial velocities 
(km~s$^{-1}$)  labeled in each panel. The spatial scale is given in one of the
panels. Fluxes have been displayed in a logarithmic scale, in  units of
$10^{-16}$~erg~s$^{-1}$~cm$^{-2}$.
\label{cha}}
\end{figure}

\begin{figure}
\centering
\includegraphics[width=\hsize]{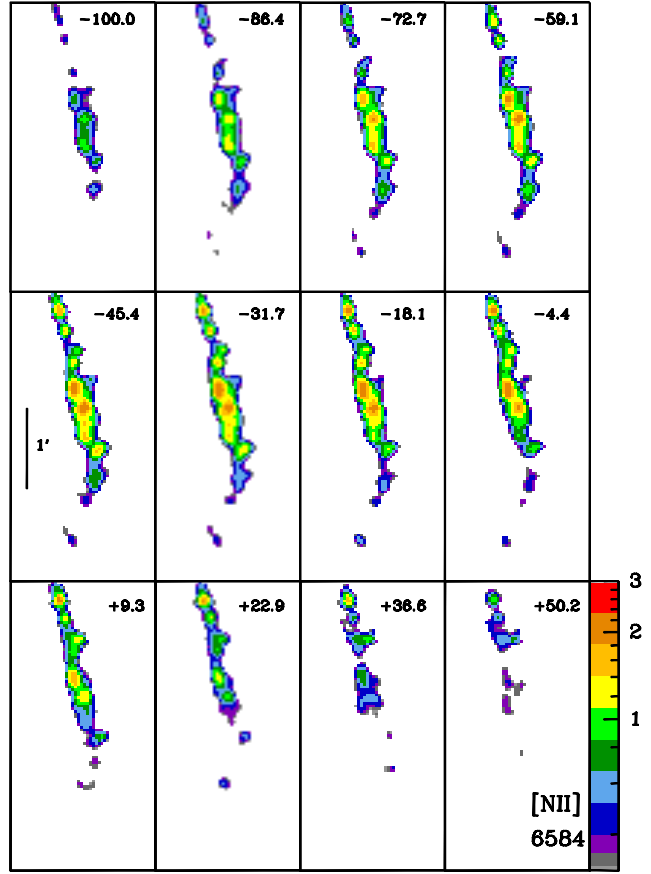}
\caption{
Same as Fig.\ \ref{cha}, but for the \nii~6584~\AA\  emission line.
 \label{cn2}}
\end{figure}

\begin{figure}
\centering
\includegraphics[width=1.05\hsize]{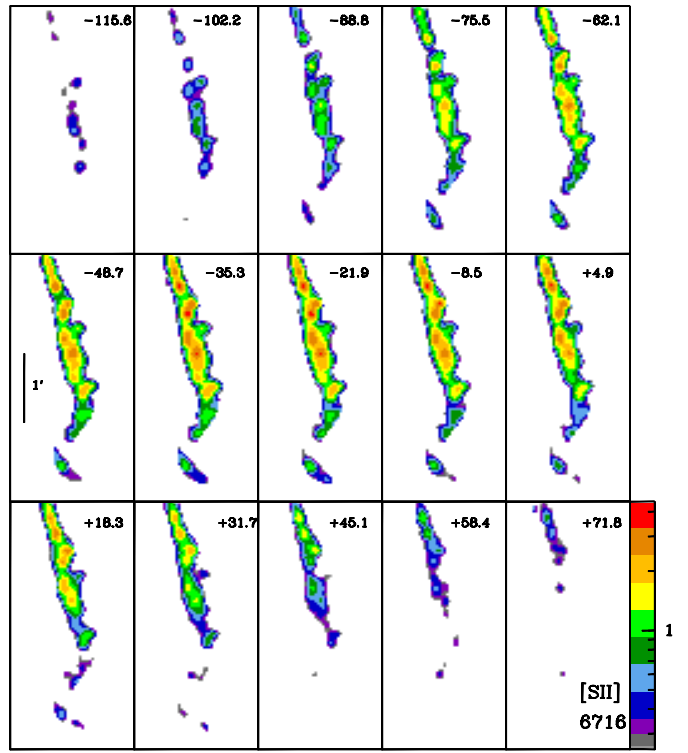}
\caption{
Same as Fig.\ \ref{cha}, but for the \sii~6716~\AA\  emission line.
\label{cs}}
\end{figure}

\subsubsection{Kinematics behaviour of physical conditions: channel maps}

A more detailed sampling of the gas kinematics was obtained by slicing the
datacube into a set of velocity channels. Each slice corresponds to a constant
wavelength bin of 0.3 \AA\ (\ie\ a velocity bin of $\sim~$14~\kms: 13.7~\kms\
for H$\alpha$ and \nii, and 13.4~\kms\ for \sii). Hence, we have obtained
the channel maps for each of the emission lines within the observed wavelength
range, namely H$\alpha$, \nii\ $\lambda$ 6584 \AA\ and \sii\
$\lambda$ 6716 \AA\ , as illustrated in Figs. \  \ref{cha}, \ref{cn2} 
and \ref{cs}, respectively. 
Moreover, we also analysed the behaviour of the excitation
and density as a function of the velocity.

The emission detected above a 3$\sigma$ SNR level 
spreads over different velocity ranges,
depending on the line. Emission in  H$\alpha$ was detected over a wider
velocity range ($\simeq$~200~\kms, from --135 to +75~\kms) than in \sii\  
($\simeq$~180~\kms, from --115 to +70~\kms) and in \nii\ ($\simeq$~150~\kms, 
from --100 to +50~\kms).  Note, however that the morphology of the emission 
does not significantly differ among the three lines when channels of similar
velocities are compared.

H$\alpha$ emission covering the whole velocity range was only detected  at the
northern, more collimated flow region, from knots A to H.  For velocities ranging
from $\simeq$~--80 to --10~\kms, there is emission detection from all the  knots, the
stronger emission coming from  the $\simeq$~--50 to --23~\kms channels.  Emission
from velocities more blueshifted than $\simeq$~--90~\kms or more redshifted than
$\simeq$~+20~\kms were not detected (up to a 3$\sigma$ snr level) beyond knot Q, from
the loci where the flow trajectory appreciably changes its path.  The kinematic
behavior in the \nii\ emission is  similar to  that found in  H$\alpha$, although the
velocity range with  emission detection is different.  In this sense,  the \nii\
emission arising from most of the  knots covers a range  from $\simeq$~--60 to
+10~\kms, the stronger emission coming from the $\simeq$~--45 to --18~\kms channels.
\nii\ emission was not detected from  the southern knots, beyond knot Q, with the
exception of some hint ($\leq 2 \sigma$ snr level) at $\simeq$~--18~\kms around knot
R. Note in addition the lack of \nii\  emission from knots A to C for velocities more
blueshifted than $\simeq$~--90~\kms . Finally, we detected \sii\ emission from the
whole flow covering a range from $\simeq$~--90 to +18~\kms, the stronger emission
coming from the $\simeq$~--35 to --8~\kms channels. As in the case of H$\alpha$ and
\nii\ , there are neither emission  more blueshifted than  $\simeq$~--90~\kms nor
more redshifted than $\simeq$~+20~\kms  detected beyond knot Q.

\begin{figure}
\centering
\includegraphics[width=\hsize]{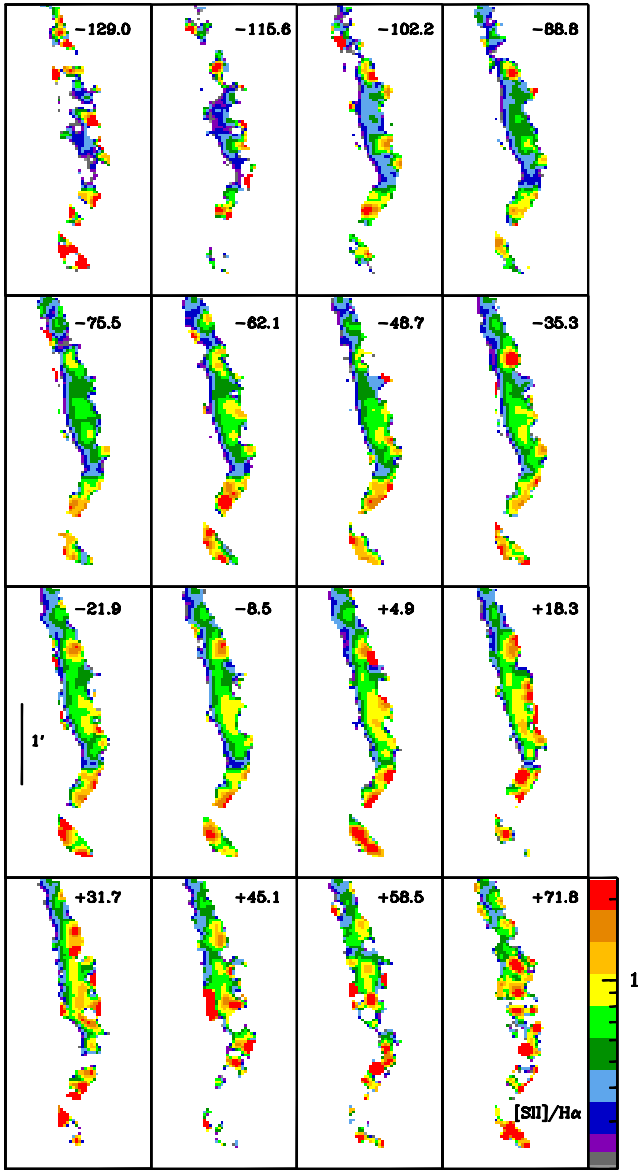}
\caption{The \sii/H$\alpha$ line-ration maps, tracing the gas excitation, 
derived from the channel maps at the \vlsr\ labeled in the corresponding panel.
\label{csh}}
\end{figure}

\begin{figure}
\centering
\includegraphics[width=1.03\hsize]{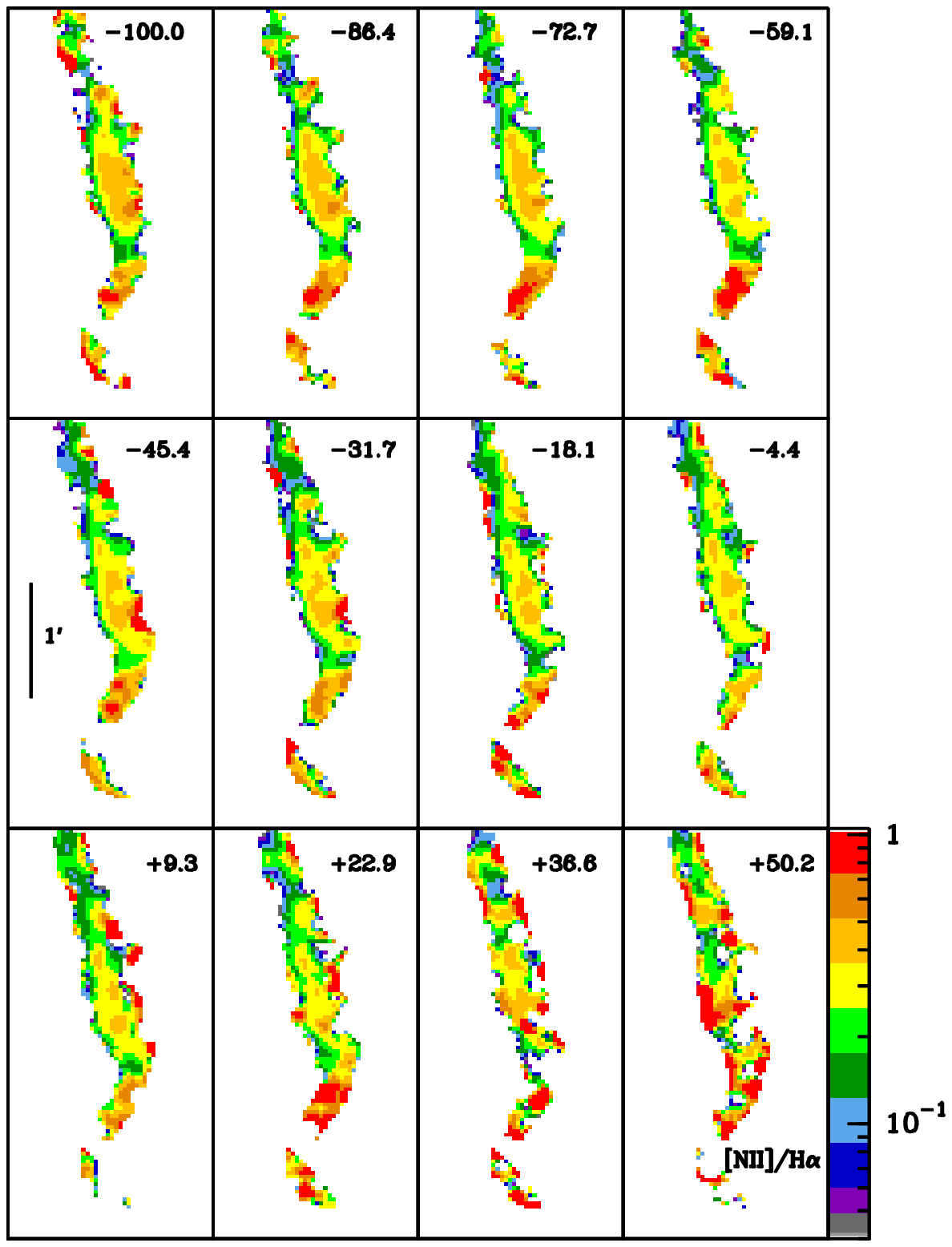}
\caption{
Same as Fig.\ \ref{csh}, but for the \nii/H$\alpha$ line-ration maps.
\label{cnh}}
\end{figure}

Line-ratio channel maps were created  from the IFU data (as mentioned in \S 3.1.2 for the
integrated line-ratio maps) to analyse the excitation
and  density as a function of the velocity. The estimated uncertainties  are
better than 25\% in the \nii 6584/H$\alpha$ and 10--15\% in  the 
\sii(6716+6731)/H$\alpha$ and \sii\ 6716/6731 maps. Figures \ref{csh} and \ref{cnh}
display the  \sii(6716+6731)/H$\alpha$ and \nii 6584/H$\alpha$  line-ratio channel
maps, tracing  the gas excitation as a function of the radial velocity.

The line-ratio maps for each velocity channel follow 
the spatial behaviour of the excitation that was already found in the
integrated  line-ratio maps (Fig.\ \ref{cfi}). The gas excitation is higher (\ie\
lower \sii/H$\alpha$ line-ratio values) at  the northern knots
(A--D), and becomes lower  (\ie\ the \sii/H$\alpha$ line-ratio values increase)
beyond knot E, along  the region where the flow cross section widens.  Furthermore,
the trend found for this region (a decreasing  excitation moving across the jet
beam, from the eastern jet side to the west) is now more clearly seen on each of
the channel map. An opposite trend  of increasing excitation from east to west is
outlined at the curved,  terminal S region from the --45~\kms~$<$\vlsr$<$+5~\kms
channels, where the signal is large enough to evaluate the ratio with confidence.

In addition, the line-ratio channel maps indicate that the gas excitation varies
depending on the gas kinematics, namely higher excitation (\ie\ lower \sii/H$\alpha$
line ratios) for the higher velocities (in absolute value, both blueshifted and
redshifted) and lower excitation (higher line ratios) for lower (in absolute value)
velocities. The differences are found for most of the knots. For example, for knot C,
at the position of its peak intensity, we measured a \sii/H$\alpha$ ratio $\leq$~0.2
for   \vlsr~$\leq$~--90~\kms\ and \vlsr~$\geq$~+30~\kms,  while a ratio of $\geq$~0.5
is found for  --50~\kms~$<$~\vlsr~$<$~+20~\kms.  The same behaviour appears for
low-excited knots:  at the peak intensity position of knot M, the \sii/H$\alpha$
ratio is  $\geq$~1.1 for --30~\kms~$<$~\vlsr~$<$~+20~\kms, being $\leq$~0.5 for
\vlsr~$\leq$~--90~\kms\ and \vlsr~$\geq$~+30~\kms. 

\begin{figure}
\centering
\includegraphics[width=1.01\hsize]{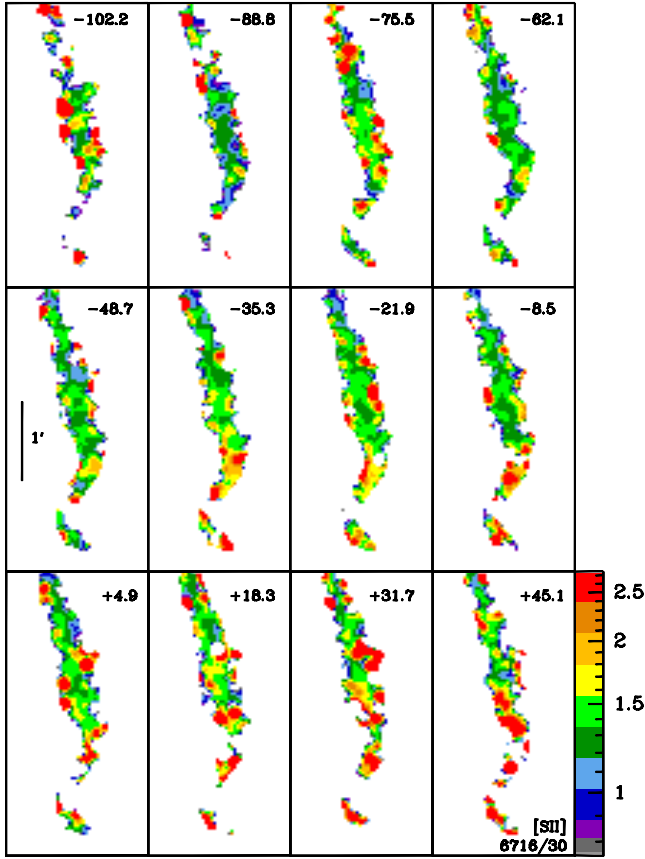}
\caption{The \sii 6716/6731 line-ratio maps, tracing the electron density,
derived from the channel maps at the \vlsr\ labeled in the corresponding panel.
\label{css}}
\end{figure}

The \sii~6716/6730 line-ratio channel maps (Fig.\ \ref{css}) trace the 
electron density of the emitting gas as a function of the radial velocity. In general, the spatial
distribution of the density follows the same trends that were  found
from the integrated \sii~6716/6730 line-ratio (Fig.\ \ref{cfi}).  In this sense,
density decreases from north to south along the jet, but  with some exception,
\eg\ around knot E, where  density is the largest. Variations on the  electron
density also depend on the gas kinematics. For a given knot, the largest 
electron density corresponds to the emission moving at lower (in absolute value)
velocities (typically within the range  --25~\kms$<$\vlsr$<$+5~\kms), whereas
the density falls down to $\simeq$ 40--50\% for the emission moving at high
velocity (\vlsr $\geq$~+20~\kms and \vlsr $\leq$~--60~\kms).  Hence, the
emission from  slower gas appears denser and less excited than the faster (red
or blueshifted) emission, which is more excited and rarefied.

\section{Summary and Conclusions}

To conclude, the IFS data confirm the peculiar nature of HH~110 as compared with most of
the stellar jets already observed. The most relevant results  presented in this paper
might be summarized as follows:

\begin{itemize}   

\item  
These IFS data have allowed us to generate the first \nii\
narrow-band  image of HH~110 currently obtained, which shows a  similar jet  cross
section to the \sii\ emission. In addition, these data allow us to confirm that  the
faint filamentary emission labeled X and Y, detected to the east of knots A--C, which
probably corresponds to a ''fossil'' outflow channel, mainly should arise from
''true'' H$\alpha$ line emission. The contributions  from nearby continuum or from
the \nii\ lines, if any, should be very weak, as no emission was detected  in \nii,
and the continuum contribution was previously removed.

\item 
The line-ratio
maps derived from the integrated line intensities revealed  complex spatial
structures for the gas excitation and  electron density. Although the 
line ratios that trace gas excitation are indicative of an
intermediate/high-excitation degree for most of the emission mapped, several
regions of low-excitation emission are detected through the flow (\eg\
\sii/H$\alpha$~$\geq$~1.5 around knot H). 
In  general, the \sii\ line ratios
correspond to low electron density values ($n_\rmn{e}$~$\leq$~1000~cm$^{-3}$)
and show smooth spatial variations. However,
strong local enhancements of $n_\rmn{e}$ relative to its
surroundings were detected at several locations (\eg\ around knot E, where
the $n_\rmn{e}$ value is about three times that of the surroundings).
 
\item 

Radial velocities appear blueshifted, relative to the \vlsr\ of the cloud, the
modulus increasing with the distance to knot A, although with some appreciable  local
departures from this trend (in particular, knot K).  This systematic increase in
radial velocity should not be interpreted as  a sign of jet acceleration, since no
similar trend was derived for proper motions. In fact, the trend found for tangential
velocities goes in the opposite way (decreasing velocity as a function of distance
from knot A, with some appreciable local departures, \eg\ in knot B).  Thus, this
apparent acceleration should be better attributed to a geometric effect: the
interaction between the jet and the inhomogeneous ambient surroundings will lead to 
the deflection of the jet  and therefore, to changes in  the projection of the full spatial
velocity along the line
of sight (radial velocity) and along the plane of the sky (tangential velocity). 

\item
In general, the line profiles are broader for
H$\alpha$ than for \sii, and the FWHM in both  lines decreases from the northern
to the southern knots, although  this trend is not fully systematic along the
outflow. 

\item  We recalculated the full velocity and the  motion direction relative to the
plain of sky of the HH~110 knots using  the \sii\ radial velocities, derived from the
spectra extracted  at the position of the knots peak intensity, together with knot
proper motions, obtained from ground-based, multi-epoch \sii\ narrow-band images.  We
confirmed the general trend  of a decrease in the velocity and an increment in the
projection angle with distance from knot A, as  found in previous works  that used a
more limited knot sample.  Significant departures from the trend are found in some
knots. These knots could be tracing the loci where a stronger interaction between the
jet and the inhomogeneous medium is taking place. 

\item    

The channel maps showed that the emission spreads over different velocity
ranges, depending on the line considered. In spite of that, the maps also showed that
the morphology of the emission in each of the   velocity channels is  similar for all
the lines, the strongest emission corresponds to the velocity interval
$\simeq$--20$<$\vlsr$<$+5~\kms, where significant emission from all the knots  was
detected. In addition,  the previously unknown kinematics of the excitation and
electron density along the full spatial extent of the outflow were obtained from the
channel maps.  The degree of excitation and the electron density appear to be
anticorrelated with the velocity modulus, in the way that the emission from the gas
moving faster (either, at red or blueshifted velocities) has higher excitation and
lower density as compared to the emission from the  gas moving at slower velocities,
which appear denser and less excited.   

\item  
From the IFS spectra extracted at appropriate spaxels, we were able
to compare the physical conditions obtained from these data with  those derived from
previous long-slit and Fabry-Perot observations. The general trends followed by the
properties of the emission and even the derived values of the observables are in good
agreement at all  the commonly sampled positions.  However, the complex structure of
the HH~110 kinematics and physical conditions,  already outlined in previous works,
has been now better  characterized from this more complete IFS sampling.  These data
revealed significant variations of the kinematics and physical conditions  over short
distances that were not properly sampled  by the long-slit data.

\item 
The emission lines have asymmetric profiles. The analysis of the H$\alpha$ line
profiles performed through the line bisectors method shows that 
the departure of the profile from a
single-Gaussian shape  varies from knot to knot. We found that the shape of the
profiles seems to be related to the full velocity and the viewing angle of the 
knot.
\end{itemize}

The scenario
first outlined by \citet{Rei96} (\ie\ deflection of the HH~270 jet by a dense
molecular cloud) and later modeled by \citet{Rag02} gives a reasonable
explanation on the origin of the HH~110 jet and offers a good
qualitative fitting
for a set of properties observed in the jet  behaviour (\eg\  the strong lack of
symmetry, relative to the outflow axis, of the proper motions), but fail
in reproducing in more detail the kinematics and some other properties through
the outflow. In particular, as  was already discussed by \citet{Lop05}, the
jet/cloud collision models of \citet{Rag02} did not predict the deceleration of
the full spatial velocity along the outflow that is derived from imaging,
long-slit and IFS data.  

A variability in the ejection outflow velocity (\ie\ pulsed jet models) and/or 
an anomalously strong interaction between the outflow and the inhomogeneous
environment were then suggested as  plausible alternatives giving rise to such
kinematic behaviour. Interestingly, new modeling that incorporates these
mechanisms has been recently published.   \citet{Yir08} and \citet{Yir09}
performed more complex, two-dimensional hydrodynamical simulations of jets that
propagate through a small-scale inhomogeneous environment (\eg\ clumps with a
size smaller  than the jet beam). Their simulations predict that the collisions
between slow clumps overtaken by faster ones should produce shock structures
slightly smaller than the jet beam and displaced from the jet axis, like what is
observed at some locations along HH~110. These simulations also predict a
complex evolution of these shock structures. In particular,  properties such as
excitation and velocity profiles across the jet beam, are far more complex than
expected from an unperturbed clump, or from the internal working
surfaces arising from  current models of a pulsed jet. 

Further insights on the behaviour of deflected supersonic jets were recently obtained
by \citet{Har09} from laboratory  experiments. The experiments show how the
morphology of the contact discontinuity between the jet and the obstacle develops  a
complex structure of cavities that gives rise to the entrainment of clumps of
material from the obstacle into the flow. The experiments also succeeded in
reproducing the morphology of some filaments such as those observed in high-spatial
resolution (HST)  images of HH~110, although the observed dynamics in these filamentary
structures were not properly reproduced by the experiments. According to the results
of these laboratory experiments, as compared to their high-resolution, long-slit
data, the authors proposed that the best model for HH~110 still is that of a pulsed
(from a varying driving source) jet interacting with a dense molecular clump.  The
results derived from our IFS observations that can be compared  with those found from
these authors (\eg\ the velocity behaviour through the flow and the line profiles at
selected positions)  appear well consistent, thus giving further support to models
that involve deflection of a pulsed jet propagating through an inhomogeneous ambient
medium.

The IFS data here presented give a full spatial coverage in the H$\alpha$, \nii\ and \sii\
emission lines of the singular outflow HH~110. They have allowed us to explore for
the first time the whole spatial distribution of the physical conditions and its
relationship with the kinematics of the jet emission.
We would like to point out that there are very few IFS observations of stellar jets with a
wide spatial coverage of the outflow in several emission lines (as \eg\ HH~34 by
\citealp{Be07}). Such observations are highly necessary to understand the behaviour of the
physical conditions as a function of the kinematics, as well as to explore whether this
behaviour varies (as in HH~34) or not (as in HH~1) with the distance from the
exciting source  (see \citealp{Gar09}). Unfortunately, the paucity of available data
prevents  to establish a general picture on these topics.

Because of the rather chaotic behaviour of HH~110 outlined in previous works, as
derived from more limited narrow-band  imaging and long-slit spectroscopic data,
the IFS observations discussed here are particularly useful 
for characterizing the properties of
the whole outflow. Hence, a more realistic picture  has arisen, suitable for
designing   new state-of-the-art simulations to match the HH~110 scenario.  Note that
IFS data give much more detailed  information  on the spatial distribution of
excitation as a function of velocity  than  provided by current model simulations or
laboratory jet experiments.  These data could provide valuable clues  to constrain
the space parameters in future theoretical works, which are necessary to understand 
the origin, structure and dynamics of HH~110.

\section*{Acknowledgments}

R.E., R.L., and A.R. was supported by the Spanish  
MICINN grant AYA2008-06189-C03-01.
BG-L thanks the support
from the Ram\'on y Cajal program by the Spanish Ministerio de Educaci\'on y
Ciencia, and the Spanish Plan Nacional de Astronom\'\i a programs AYA2006-13682,
AYA2009-12903.
SFS thanks the Spanish Plan Nacional de Astronom\'\i a program
AYA2005-09413-C02-02, of the Spanish Ministery of Education and Science
and  the Plan Andaluz de Investigaci\'on of Junta de Andaluc\'{\i}a as
research  group FQM322.
We thank A. Eff-Darwich for his useful help with the manuscript. 
R.L. acknowledges the hospitality of the Instituto de Astrof\'{\i}sica de
Canarias, where part of this work was done.

\bsp

\label{lastpage}


\begin{thebibliography}{}

\bibitem[\protect\citeauthoryear{Beck et al.}{2007}]{Be07}
Beck, T.L., Riera, A., Raga, A.C. \& Reipurth, B., 2007, AJ, 133, 1211.

\bibitem[\protect\citeauthoryear{Choi}{2001}]{Cho01}
Choi, M., 2001, ApJ, 550, 817.

\bibitem[\protect\citeauthoryear{Dall et al.}{2006}]{Dal06}
Dall, T. H.,  Santos,  N. C.,  Arentoft,  T., Bedding,  T. R. \&  
Kjeldsen, H., 2006, A\&A, 454, 341.

\bibitem[\protect\citeauthoryear{Garc\'{\i}a-L\'opez et al.\ }{2009}]{Gar09}
Garc\'{\i}a-L\'opez, R., Nisini, B., Giannini, T., Eisloeffel, J., Bacciotti, F. 
\& Podio, L., 2009, A\&A, 487, 1019.

\bibitem[\protect\citeauthoryear{Garc\'{\i}a-Lorenzo, Acosta-Pulido \&
Megias-Fern\'andez}{2002}]{Ga02}
Garc\'{\i}a-Lorenzo B.,  Acosta-Pulido J.,  Megias-Fern\'andez E., 2002, 
in ASP Conf. Ser. 282, Galaxies: The Third Dimension, ed.\ M. Rosado, L.
Binette, \& L. Arias (San Francisco: ASP), 501

\bibitem[\protect\citeauthoryear{Hartigan, Raymond \& Hartmann}{1987}]{Har87}
Hartigan, P., Raymond, J., Hartmann, L., 1987, ApJ, 316, 323.

\bibitem[\protect\citeauthoryear{Hartigan et al.}{2009}]{Har09}
Hartigan, P., Foster, J.M., Wilde, B.H., Coker, R.F., Rosen, P.A., Hansen, J.F.,
Blue, B.E., Williams, R.J.R., Carver, R., Frank, A., 2009, ApJ, 705, 1073.

\bibitem[\protect\citeauthoryear{Kelz et al.}{2006}]{Ke06} 
Kelz A., Verheijen M.A.W., Roth M.M. et al., 2006,  PASP, 118, 129 

\bibitem[\protect\citeauthoryear{L\'opez et al.}{2005}]{Lop05} 
L\'opez R., Estalella R., Raga, A.C., 
Riera A., Reipurth, B., Heathcote, S.R.,  2005, A\&A, 432, 567.

\bibitem[\protect\citeauthoryear{Mart\'{\i}nez-Fiorenzano}{2008}]{Mar08}
Mart\'{\i}nez-Fiorenzano, A.F., 2008, 
in {\it Precision Spectroscopy in
Astrophysics}, Proceedings of the ESO/Lisbon/Aveiro Conference held in
Aveiro, Portugal, 11-15 September 2006. Edited by N.C. Santos, L. Pasquini,
A.C.M. Correia, and M. Romanielleo. Garching, Germany, 2008 pp. 143-144.

\bibitem[\protect\citeauthoryear{Queloz et al.}{2001}]{Que01}
Queloz, D., Henry, G. W, Sivan, J. P., Baliunas, S. L., Beuzit, J. L.,
Donahue, R.A., Mayor, M., Naef D., Perrier, C.  \& Udry. S., 2001, A\&A, 379,
279.


\bibitem[\protect\citeauthoryear{Raga et al.}{1996}]{Rag96} 
Raga, A.C., B\"ohm, K.-H. \& Cant\'o, J. 1996, REvMexAA, 32, 161.

\bibitem[\protect\citeauthoryear{Raga et al.}{2002}]{Rag02} 
Raga, A.C., de Gouveia Dal Pino, E.M., Noriega-Crespo, A., Mininni, P.D., \&
Vel\'azquez, P. 2002, A\&A, 392, 267.

\bibitem[\protect\citeauthoryear{Reipurth \& Olberg}{1991}]{Rei91} 
Reipurth, B., Olberg, M., 1991, A\&A, 246, 535.

\bibitem[\protect\citeauthoryear{Reipurth, Raga \& Heathcote}{1996}]{Rei96} 
Reipurth, B., Raga, A.C., Heathcote, S., 1996 A\&A, 311, 989.


\bibitem[\protect\citeauthoryear{Riera et al}{2003a}]{Rie03a}
Riera, A., L\'opez, R., Raga, A.C., Estalella, R., Anglada, G., 2003a, A\&A,
400,213.


\bibitem[\protect\citeauthoryear{Riera et al}{2003b}]{Rie03b}
Riera, A., Raga, A.C., Reipurth, B., Amram, P, Boulesteix, J., Cant\'o, J.,
Toledano, O., 2003b, AJ, 126, 327.

\bibitem[\protect\citeauthoryear{Roth et al.}{2005}]{Rot05}
Roth M.~M., Kelz A., Fechner T., Hahn T., Bauer S.-M., Becker T., B{\"o}hm P.,
Christensen L., Dionies F., Paschke J., Popow E., Wolter D., Schmoll J., Laux
U.,  Altmann W., 2005, PASP, 117, 620

\bibitem[\protect\citeauthoryear{S\'anchez}{2004}]{Sa04} 
S\'anchez, S.~F.\ 2004, AN, 325, 167 

\bibitem[\protect\citeauthoryear{S\'anchez}{2006}]{Sa06} 
S\'anchez, S.~F.\ 2006, AN, 327, 850

\bibitem[\protect\citeauthoryear{S\'anchez et al.}{2007}]{San07} 
S\'anchez, S.~F., Cardiel, N., Verheijen, M.A.W., Mart\'{\i}-Gordon, D.,
Vilchez, J.M., Alves, J., 2007, A\&A, 465, 207.

\bibitem[\protect\citeauthoryear{Schultz, Burton \& Brand}{2005}]{Sch05}
Schultz, A.S.B., Burton, M.G., Brand, P.W.J.L., 2005, MNRAS, 358, 1195. 
 
 \bibitem[\protect\citeauthoryear{Sep\'ulveda et al.}{2010}]{Sep10}
Sep\'ulveda, I., Anglada, G., Estalella, R., L\'opez, R., Girart, J.M., Yang,
J., 2010, A\&A submitted.

\bibitem[\protect\citeauthoryear{Yirak et al.}{2008}]{Yir08}
Yirak, K., Frank, A., Cunningham, A.J., Mitran, S., 2008, Apj, 672, 996.

\bibitem[\protect\citeauthoryear{Yirak et al.}{2009}]{Yir09}
Yirak, K., Frank, A., Cunningham, A.J., Mitran, S., 2009, ApJ, 695, 1005.


\end{thebibliography}
\end{document}